\documentclass[10pt,twoside]{hsqcd}
\usepackage{amsmath}
\usepackage{graphicx}

\begin{document}

\title{WHAT IS THE REAL MEANING \\
       OF THE FROISSART THEOREM?
\thanks{This work is partly
supported by the Russian State Grant
No. RSGSS-65751.2010.2.}}

\author{Ya.~AZIMOV\\ \\
St. Petersburg Nuclear Physics Institute \\
Gatchina, Russia\\
E-mail:  azimov@thd.pnpi.spb.ru}

\maketitle

\begin{abstract}
\noindent The Froissart bounds for amplitudes and cross sections
are explained and reconsidered to clarify the role of different
assumptions. It is the set of physical conditions of unitarity and 
of no massless exchanges, together with mathematical properties
of the Legendre functions, that implies much softer high-energy
asymptotics for elastic amplitudes at physical angles as compared to 
the same amplitudes at nonphysical angles. The canonical log-squared
boundary for $\sigma_{tot}$ appears only under the additional
hypothesis that the amplitude at any nonphysical angle cannot grow
faster than some power of energy. The Froissart results are further
shown to admit some reinforcement. Comparison of the familiar and
new Froissart-like restrictions with the existing data on
$\sigma_{tot}$ and diffraction slope at all available energies
(including LHC) does not allow yet to unambiguously determine the
asymptotic behavior of $\sigma_{tot}$, but shows that its current
increase cannot be saturated (\textit{i.e.}, maximally rapid).

\end{abstract}



\markboth{\large \sl  Ya.~Azimov
\hspace*{2cm} HSQCD 2012} {\large \sl \hspace*{1cm}
What is the real meaning
 of the Froissart theorem ? }

\section{What is the Froissart theorem?}

The Froissart theorem (or Froissart bound) is known since 1961,
after publication of the paper~\cite{fr}. Its main statement says
that the total cross section of two-hadron interaction cannot grow
with energy faster than $\log^2E$. Moreover, according to the common
opinion frequently repeated in the literature, violation of this
limit would mean violation of unitarity.

All preLHC experimental data on total cross sections look consistent
with such expectation (see \textit{e.g.}, Ref.~\cite{block}). This is
especially impressive for the nucleon--(anti)nucleon scattering, where
Tevatron and cosmic rays provided much higher energies than for any
other hadron pairs (note, however, large uncertainties of the cosmic
ray data). Nevertheless, the same data admit also ``heretic'' descriptions,
with faster energy growth (\textit{e.g.}, fit~\cite{land} has a power
of energy in asymptotics).

It is reasonable, therefore, to reconsider the theoretical background
of the Froissart theorem to answer the following questions:
\begin{itemize}
\item  Why, at all, quantum theory may provide any bounds (the Froissart
bounds  (F.b.) in particular) for energy increase of amplitudes and
cross sections?
\item  How does the specific bound  $\sim\log^2E$  arise?
\end{itemize}
\section{Main steps of Froissart's construction}

Let us briefly recall the derivation of the Froissart theorem. In his
paper~\cite{fr}, Froissart used the following assumptions: a) unitarity
condition; b) strong interactions as an object for discussion; c) elastic
amplitudes satisfy the Mandelstam (double-spectral) representation or,
at least, the single dispersion relation in the momentum transfer; d) total
cross sections grow with energy.

From these assumptions Froissart deduced several bounds:
a) for forward (or backward) amplitudes ($<s\log^2s$, where $s$ is
the c.m. energy squared); b) for total cross sections, as a consequence
of the optical theorem ($<\log^2s$); c) for fixed-angle amplitudes
($<s^{3/4}\log^{3/2}/\sin^{1/2}\theta$; for simplicity, constant factors
are omitted in all inequalities).

Derivation of those bounds begins with the familiar partial-wave
decomposition
\begin{equation}
A(s,t)\ =\ \sum_{l=0}^{\infty} (2l+1)\,a_l(s)\,P_l(z)\,,
\label{pwa}
\end{equation}
where $t=2k^2(-1+z),~~ z=\cos\theta\,$. Due to unitarity, $\,|a_l(s)|<1\,$.
One more inequality is generated by the dispersion relation. It provides the
Gribov--Froissart representation for $a_l$~\cite{fr, gr}, due to which
$|a_l|<B(s)\exp(-l\,\alpha_0)$ at large $l\,$. The value of $\alpha_0$ is
related to the nearest $t$-channel singularity $t_0=2k^2(-1+\cosh\alpha_0)\,$;
at high energy $\alpha_0^2\sim s^{-1}$. We have thus two boundaries for $|a_l|$,
which intersect at $l=L\,$. For $\sigma_{tot}$ to grow, $L$ should grow as well.
It appears then that only sum with $l\leq L$ is important at high energies, and
the asymptotic behavior of the amplitude is completely determined by the
behavior of $L\,$. For the forward/backward scattering amplitude, where
$|P_l(\cos\theta)|=|P_l(\pm1)|=1$,
\begin{equation}
|A(s,0)|\ <\ \sum_{l=0}^L \,(2l+1)\sim L^2\,,
\label{fwda}
\end{equation}
while for the fixed-angle amplitude, with
$|P_l(\cos\theta)|<\sqrt{2/(\pi l\sin\theta)}$,
\begin{equation}
|A(s,z)|\ <\ \sum_{l=0}^L\,(2l+1)/(l\sin\theta)^{1/2}\ \sim\
L^{3/2}/(\sin\theta)^{1/2}\, .
\label{faa}
\end{equation}
The value of $L$ is determined by the relation $1=B(s)\,\exp(-L\,\alpha_0)\,$.
Dispersion relation implies $B(s)\sim (s/s_0)^N$ with fixed $s_0$, and
$L\sim s^{1/2}\log(s/s_0)$. This directly leads to Froissart's results.

\section{Modified approach to Froissart's bounds}

Froissart's approach was recently reconsidered and modified~\cite{az}. The new
approach uses the following assumptions: a) unitarity condition, just as before;
b) absence of massless particles (it is just this point that marks strong
interactions, no other assumptions about properties of interaction are used);
c) no assumptions on dispersion relations. This set of assumptions provide
bounds for elastic amplitudes and total cross sections, which generalize
Froissart's ones.
\subsection{Main steps of the modified approach}

As before, we have the unitarity bound $|a_l|<1$. To obtain another one,
we can transform the usual relations
\begin{equation}
a_l(s)\ =\ \frac{\pi k}{2\sqrt{s}}\int\limits_{-1}^{+1}
A(s,\cos\theta^{'})\,P_l(\cos\theta^{'})\,
d(\cos\theta^{'})\ =\ -\frac{i k}{2\sqrt{s}}\oint A(s,
z')\,Q_l(z')\,dz'\,,
\label{pw-am}
\end{equation}
where $Q_l$ is the Legendre function of the 2nd kind. The closed integration
contour runs anticlockwise around the cut of $Q_l(z')$ between $-1$ and $+1$
(it is the only cut of $Q_l(z')$ at positive integer $l$). We can choose the
contour to be an ellipse with $z'\equiv\cosh(\alpha+i\phi'),~\alpha=
\mathrm{const}>0,~ -i\,dz'=\sin(\alpha+i\phi')\,d\phi'\,$. The contour may be
blown up, until it touches the nearest singularity with $\alpha=\alpha_0>0$
(in the physical region $\alpha=0$). Then, at large $l$, we can apply the
inequality (equivalent to one used by Froissart)
\begin{equation}
|\sin(\alpha+i\phi')\,Q_l(z')|\ <\ e^{-\alpha(l+1/2)}\,\sqrt{\frac{\pi}{2}\,
\cosh\alpha} \label{q1}
\end{equation}
(the right-hand side independent of $\phi'\,$!), to obtain $|a_l|<
B_0(s)\,\exp(-l\alpha_0)\,$, where $B_0(s)$ is determined by the nonphysical
amplitudes $A(s,z')$ on the integration contour. Again, we obtain ``critical''
value $L$, satisfying the relation
\begin{equation}
e^{\alpha_0\,L}\ =\ \frac12\,B_0(s)\,,
\label{L0}
\end{equation}
and the whole set of Froissart's bounds in terms of $L$. Specific
form of $s$-asymptotics depends on asymptotics of $B_0(s)$. Original Froissart's
results~\cite{fr} are reproduced if $B_0(s)$ grows as a power of $s$.

Though $B_0(s)$ and $B(s)$ look differently, they both are determined by the
amplitudes $A(s,t)$ with nonphysical values of $t$ and should, thus, have similar
high-$s$ behavior. The power dependence of $B(s)$ was motivated by dispersion
relations which, however, have never been generally proved. The modified
approach does not need any dispersion relation, and asymptotics of $B_0(s)$
becomes a completely independent assumption, not related neither to unitarity
nor to analyticity. Moreover, as explained in Ref.~\cite{az}, phenomenological
linearity of hadronic Regge trajectories gives an indirect evidence for $B_0(s)$
increasing faster than any power of $s$, {\em i.e.}, for $\sigma_{tot}$
increasing faster than $\log^2s$.

In any case, high-energy asymptotics for physical amplitudes (and cross sections)
is much softer than that for nonphysical amplitudes. It is a very
general consequence of the assumptions described above.

\subsection{Enhancing the Froissart bounds}

The inequality (\ref{q1}) is rather loose. Its more exact and stricter form is
\begin{equation}
|\sin(\alpha+i\phi')\,Q_l(z')|\ <\ e^{-\alpha(l+1/2)}\,
\sqrt{\frac{\pi}{2l}\,\cosh\alpha}\,.
\label{q2}
\end{equation}
This leads to a different value of $L$, as determined by the relation
\begin{equation}
e^{\alpha_0\,L}\,\sqrt{L}\ =\ \frac12\,B_0(s)\,.
\label{L1}
\end{equation}
The new value is smaller than the previous one, from Eq.(\ref{L0}), and leads
to stricter boundaries for physical amplitudes and cross sections. For example,
if $B_0(s)\sim s^N$ (as usually assumed), then $\sigma_{tot}$ cannot increase
as $\sim\log^2(s/s_0)$ with a fixed scale $s_0$ (as usually stated). Instead,
the scale $s_0$ should itself increase with energy (as some power of $\log{s}$).

\subsection{New Froissart-like inequalities}

Originally, Froissart~\cite{fr} obtained restrictions for the forward/backward
amplitudes ($<L^2$) and for the fixed-angle ones ($<L^{3/2}/{\sin^{1/2}\theta}$).
Analysis of Ref.\,\cite{az} gave also inequalities for fixed-$t$ cases,
not considered by Froissart. In particular, for physical (negative)
values of $t$ it gives $|A(s,t)|<L^{3/2}(s/|t|)^{1/4}$ (for all
inequalities here we omit constant factors).

Of course, all those inequalities provide restrictions, not prescriptions. If,
nevertheless, the forward amplitude (as well as $\sigma_{tot}$) is saturated,
\textit{i.e.}, grows with energy as fast as possible, then it definitely grows
faster than any physical fixed-$t$ amplitude. This necessitates existence
of the shrinking diffraction peak. The slope $b$ of this peak should grow just
as $\sigma_{tot}$ or even faster~\cite{az}.

Therefore, at high energies the ratio $\,\sigma_{tot}/b\,$ cannot increase, it
should either decrease or be a constant. It is worth to emphasize that this
conclusion is very general: it is the result of the above mentioned assumptions
of unitarity and absence of massless particles (sure, both are true for the
strong interactions), appended by the not evident assumption of saturated total
cross section.

This result has an interesting physical meaning. If one considers the
hadron-hadron high-energy scattering as diffraction on a screen, then the slope
$b$ is proportional to the screen area, while the ratio $\,\sigma_{tot}/b\,$ is
proportional to the average blackness of the screen. Thus, in the high-energy
asymptotics the average blackness should be either constant or decreasing. The
former case might, in particular, correspond to completely black hadrons (as
usually assumed).

The latter case means that the screen area increases faster than
$\,\sigma_{tot}\,$ and the screen, in average, becomes more transparent. This
seems paradoxical, but for comparison we can recall the case of a constant
$\,\sigma_{tot}\,$, where a constant slope in strong interaction scattering
would contradict analyticity and $t$-channel unitarity~\cite{gr2}. The
contradiction may be overcome by the Reggeon description (with the unit
intercept), in which the target has an infinitely increasing radius (and
area) and becomes more and more transparent. This case corresponds to
$\,\sigma_{tot}/b\sim 1/\log{s}\,$ and shows that a hadron at high energies
may tend to become completely transparent, contrary to intuitive expectations
for strong interactions.

\section{Current experimental situation}

The LHC measurements of total and elastic cross sections~\cite{tot} are a
great progress. Their results agree quite well with the log-squared behavior of
$\,\sigma_{tot}$, but they do not provide yet an unambiguous answer and still
may be described ``heretically'', with power increase~\cite{land2}.

More definite (and intriguing) is the energy behavior of the ratio
$\,\sigma_{tot}/b\,$ shown in Fig.\,1. At energies from 10\,GeV to
100\,GeV this ratio is nearly constant (or slightly decreasing).
However, it definitely increases when going to the LHC energy.
\begin{figure}[!thb]
\vspace*{7.0cm}
\begin{center}
\includegraphics{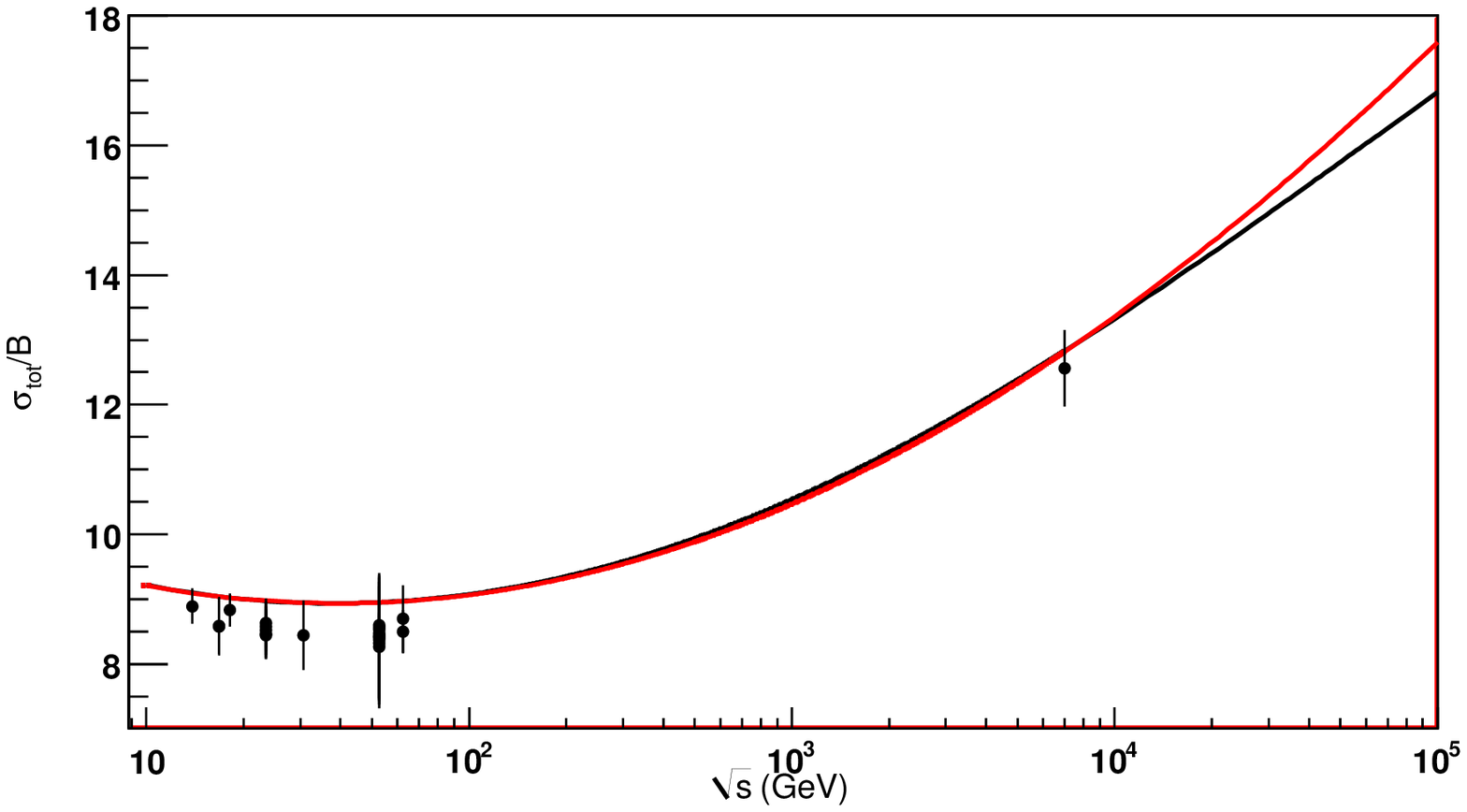}
\caption[*]{\small The energy dependence of the ratio $\sigma_{tot}/b\,$
(the figure is taken from Ref.\,\cite{fag-men}). }
\end{center}
\vspace*{-0.5cm}
\end{figure}

According to the above analysis, such a result, if confirmed, means
that the observed growth of $\sigma_{tot}$ cannot be saturated.

\section{Conclusions}

Let us summarize results of the analysis~\cite{az}.
\begin{itemize}

\item The very general result, which is \textit{the real meaning} of the
   Froissart theorem, is the much softer energy growth for physical
   amplitudes \textit{vs}. nonphysical ones. This is based on the physical
   assumptions of unitarity and absence of massless particles together with
   mathematical properties of the Legendre functions.

\item Particular form  of high-energy asymptotics of $\sigma_{tot}$  is,
   theoretically, an open question. Commonly believed  log-squared one is
   related  to an additional suggestion (\textit{never justified}) of
   not-stronger-than-power growth for amplitude(s) in any nonphysical
   configurations. Violation of the log-squared behavior would not violate
   unitarity; it would contradict only to the additional asymptotic assumption.

\item Familiar Froissart bounds may be enhanced; under the familiar assumptions,
   $\sigma_{tot}$ can not grow faster than $\ln^2(s/s_0)$ with a growing scale
   $s_0$, \textit{i.e.}, must grow \textit{slower} than canonically assumed.

\item There are indirect arguments for cross sections to grow with energy
   \textit{faster} than the ``canonical'' log-squared one.

\item The observed relation between $\sigma_{tot}$ and the diffraction slope
   $b$ gives evidence that the presently observed energy growth is
   \textit{not saturated}.

\item Further studies, both theoretical and experimental, are necessary.

\end{itemize}


\end{document}